\documentclass[apl,twocolumn,showpacs,superscriptaddress]{revtex4}
\usepackage{amsmath,amssymb,graphicx,epstopdf,color,bm}

\begin{document}

\title{An exciton-polariton mediated all-optical router}

\author{H. Flayac}
\affiliation{Institute of Theoretical Physics, Ecole Polytechnique F\'ed\'erale de Lausanne (EPFL), CH-1015 Lausanne, Switzerland}
\email{hugo.flayac@epfl.ch}

\author{I. G. Savenko}
\affiliation{COMP CoE at the Department of Applied Physics and Low Temperature Laboratory
(OVLL), Aalto University School of science, P.O. Box 13500, FI-00076 Aalto, Finland}
\email{Ivan.Savenko@aalto.fi}

\begin{abstract}
We propose an all-optical nonlinear router based on a double barrier gate connected to periodically modulated guides. A semiconductor microcavity is driven nonresonantly in-between the barriers to form an exciton-polariton condensate on a discrete state that is subject to the exciton blueshift. The subsequent coherent optical signal is allowed to propagate through a guide provided that the condensate energy is resonant with a miniband or is blocked if it faces a gap. While a symmetric sample operates as an optical switch, its asymmetric counterpart embodies a router turned to be polarization selective under applied magnetic field.
\end{abstract}


\maketitle

%_________________________________________
%--------------------------------------------------------------

%\emph{Introduction.---}
Coherent emitters of light, all-optical transistors and valves~\cite{GaoPRB2013,BallariniNatCom2013}, sources of infrared and terahertz radiation~\cite{THzPRL1,THzPRL2}, nonlinear logic elements, optical integrated circuits or interferometers ~\cite{Imamoglu,LiewPRL2008,PavlovicPRL2009,JohnePRB2010,EOArX2013,Racquet,PolaritonDevices} -- this is a non-exhaustive queue of potential exciton-polariton based devices, some of which have already found their experimental implementation thanks to the remarkable progresses in sample growth and etching technologies. Although there still does not exist an industrial polariton-based park of equipment, this area of the condensed matter realm continues to spread and appeal versatile interest of the scientific community.

A set of distinctive but peculiar properties of polaritons engender a series of intriguing fundamental many-body effects and phenomena ranging from quasi-condensation \cite{KasprzakNature} and superfluidity \cite{AmoNature} to the Josephson effect \cite{LagoudakisJosephson,AmoJosephson} or polariton-mediated superconductivity~\cite{LaussySupercond}.
Naturally, the utmost fascinating feature of exciton polaritons - which represent exciton-photon mixtures - is the possibility to control, manipulate and monitor high-speed disorder-proof optical signals. Indeed, as it was discussed in numerous works, both the light and matter components of a polariton turn out beneficial: excitons, on the one hand, are subject to the dipole and exchange interactions, and thus serve as a control gear; confined photons, on the other hand, having small effective mass, superinduce passibility of the hybrid particles.

Furthermore, a characteristic pseudospin structure of polaritons \cite{ReviewSpin} together with the possibility of their lateral confinement~\cite{Confinement} has opened a promising field of research called "spin-optronics", based on the spin transport of bosonic particles. Obviously, in such investigations the chip architecture based on quasi one-dimensional artificial microwires conquers the attention of researchers \cite{WertzNature,LiewCircuit,TaneseGS}.
Beyond the polariton physics, on the basis of these 1D channels, it is possible to create tunneling effect-based devices which possess various nonlinear properties~\cite{Sollner1983,Mizuta1995}. Evidently, the simplest and the most popular sketch has been a double barrier structure, which allows investigating, for instance, the Coulomb blockade effect~\cite{Pasquier1993,Joyez1994}, or quantum transport of atomic Bose-Einstein condensates \cite{Carusotto2000,Paul2005}.
In the polariton case, there exist several techniques of the potential profile engineering. It can be embedded either optically~\cite{AmoPRB2010,TosiNature2012}, electrically~\cite{LaiNature2007} or mechanically, by etching the microcavity~\cite{BlochSM1997,ParaisoPRB2009}. Recently, a long-awaited experimental demonstration of the control of polariton tunneling through the double barrier structure \cite{NguyenPRL2013} was reported. 	
\begin{figure}[ht]
\includegraphics[width=0.5\textwidth,clip]{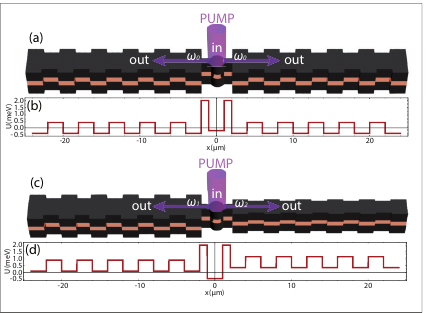}
\caption{Sketches of species of the device consisting of etched microcavities. A central input island is connected to two periodically modulated waveguides (photonic crystalls). (a) Symmetric case: the sample can behave as an optical gate. (c) asymmetric case allowing optical routing. $\omega_j$ correspond to the output frequencies. The effective confining potential patterns embedded in the structure are shown in panels (b) and (d), respectively.}
\label{Fig1}
\end{figure}

Capitalizing on these remarkable results, in this letter we propose several signal tracing effects in a setup, in which the two-barrier quantum structure determines the behavior of the localized condensate of polarized particles. It will be demonstrated, that on the basis of such an implement it turns out possible to achieve several logical setups, namely, an optical gate and a spin selective optical router.
%--------------------------------------------------------------
%--------------------------------------------------------------
%--------------------------------------------------------------

We shall describe the dynamics of a spinor polariton field, $\boldsymbol{\psi}=(\psi_+,\psi_-)^T$, through a set of spin dependent one-dimensional Gross-Pitaevskii equations coupled to an excitonic reservoir:
\begin{eqnarray}
\label{psi1}
i\hbar \frac{{\partial {\psi _ + }}}{{\partial t}} &=& \left[ \begin{array}{l}
 - \frac{{{\hbar ^2}}}{{2m}}\frac{\partial^2 }{{\partial x^2}} + U + {\alpha _1}\left( {{{\left| {{\psi _ + }} \right|}^2} + {n_R}} \right)\\
 + {\alpha _2}{\left| {{\psi _ - }} \right|^2} - \frac{{i\hbar }}{2}\left( {\Gamma  - \gamma {n_R}} \right) - \frac{{{H_z}}}{2}
\end{array} \right]{\psi _ + }\\
\label{psi2}
i\hbar \frac{{\partial {\psi _ - }}}{{\partial t}} &=& \left[ \begin{array}{l}
 - \frac{{{\hbar ^2}}}{{2m}}\frac{\partial^2 }{{\partial x^2}} + U + {\alpha _1}\left( {{{\left| {{\psi _ - }} \right|}^2} + {n_R}} \right)\\
 + {\alpha _2}{\left| {{\psi _ + }} \right|^2} - \frac{{i\hbar }}{2}\left( {\Gamma  - \gamma {n_R}} \right) + \frac{{{H_z}}}{2}
\end{array} \right]{\psi _ - }\\
\label{nR}
\frac{{\partial {n_R}}}{{\partial t}} &=& {P_R} - \left[\gamma \left({{{\left| {{\psi _ + }} \right|}^2} + {{\left| {{\psi _ - }} \right|}^2}}\right) + {\Gamma _R} \right]{n_R}
\end{eqnarray}
Here $\alpha_{1}$ and $\alpha_2$ are the parallel and antiparallel spin interaction strengths respectively, $\Gamma$ is the polariton radiative decay rate and $H_z$ is the Zeeman splitting imposed by an external magnetic field, $\mathbf{B}=B_z \mathbf{e}_z$. Eq.(\ref{nR}) is a rate equation for the dynamics of an unpolarized exciton reservoir coupled to the system, where the exchange of excitations is characterized by the decay rate, $\Gamma_R$, and the frequency, $\gamma$.

This phenomenological model \cite{WCModel} can in most cases efficiently describe a formation of a polariton condensate under nonresonant excitation \cite{WertzNature} (far away from the excitonic resonances) together with the reservoir dynamics. It should be noted, however, that while the stimulated scattering is taken into account, the phonon-mediated energy relaxation processes are neglected. They can be accounted for by a modified set of equations~\cite{LiewTransistor,SavenkoPRL2013}, though, we shall insure that the condensate forms on the lowest energy state with an initial seed. Furthermore, the ballistic propagation of polaritons was demonstrated over tens of micrometers in Ref.\cite{WertzNature} without exhibiting relaxation away from the pumping spot.

Let us consider the sample modeled in Fig.\ref{Fig1}(a) and the effective potential profile, $U(x)$, it produces within a 1D mapping [Fig.\ref{Fig1}(b)]. It consists of a central input region separated from two output waveguides by square barriers similarly to Ref.\cite{NguyenPRL2013}. However, here the operative mode is fundamentally different. Indeed, the resonant tunneling diode involves a coherent input beam in one of the guides. The signal is either blocked or allowed to tunnel through the double barrier structure by the monitoring second gate beam, for which the condensation is rather an idler or even detrimental effect. In the configuration we propose, not only the guides are periodically modulated but the input region is situated in-between the barriers where the polariton states are confined to zero dimension and bound to discreetness. The system is excited in this region by a nonresonant pump, $P_R(x)$, thus creating a polariton condensate above threshold. The latter constitutes the gate and the coherent signal which is to be routed after its potential tunneling into the guides.

The excitonic reservoir forms a potential barrier for the polaritons \cite{WertzNature} and, as a result, their discrete energy spectrum becomes blueshifted by a quantity $\mu_{\rm{res}}=\alpha_1 n_R$, where $n_R$ is the exciton density. As soon as the polariton condensate is formed, the reservoir becomes partially emptied, but the polariton-polariton interactions come into play. They blueshift the states in turn by a quantity $\mu_\pm=(\alpha_1+\alpha_2)|\psi_\pm|^2$. Hence, each of the spin components suffers a total blueshift amounting to $\mu_{\rm{res}}+\mu_\pm$ increasing at the same time the polariton emission temporal frequency.

The periodic modulation of the guides is then the key ingredient of the signal switching. Sure enough, as it has already been considered in Refs.\cite{LaiNature2007,FlayacBOs1,FlayacBOs2,TaneseGS}, it is known to form minibands in the polariton dispersion separated by gaps. Therefore, the output signal can be switched on and off depending on whether the localized state is resonant with a miniband or not owing to the state elevation.

We have considered an InAlGaAs-based alloy and thus used the following set of parameters assuming here zero detuning between the excitonic and photonic modes: The polariton mass $m=4.55\times10^{-35}$ kg; spin interaction strength: $\alpha_{1}=1.2\times10^{-3}$ meV$\cdot\mu$m and $\alpha_2=-0.1\alpha_1$; the polariton radiative decay rate $\Gamma=10^{-2}$ps$^{-1}$; the reservoir decay rate: $\Gamma_R=2.5\times10^{-3}$ps$^{-1}$ and the transfert frequency: $\gamma=5\times10^{-6}$ps$^{-1}$. Sample geometry: the central input region is characterized by 1 $\mu$m thick and 2 meV high barriers separated by 2 $\mu$m. The output guides have a modulation period $d=4$ $\mu$m and an amplitude $A_p=$0.4 meV. It forms the first and second allowed bands of the widths 0.1 meV and 0.55 meV, respectively and the gap is 0.4 meV wide. Technologically, the modulation of the guides can be achieved by metallic depositions \cite{LaiNature2007}, surface acoustic waves \cite{SAW} or a lateral etching \cite{TaneseGS}.
As shown in Fig.\ref{Fig1}(a,c), we tend to favor the latter solution modulating the photonic confinement energy, since it does not reduce the polariton lifetime and allows to delimit the periodic region with the required precision. The drawback is the fixed periodicity and amplitude of the potential. The two transverse modulation widths, $\ell_1$ and $\ell_2$, are linked to $A_P$ via
\begin{equation}\label{l1l2}
{\ell_2} = \frac{{{\ell_1}{\pi ^2}{\hbar ^2}}}{{2{\ell_1}{A_p} + {\pi ^2}{\hbar ^2}}}.
\end{equation}
Account for the nonparabolicity of the polariton dispersion and separation the excitonic and photonic fields would slightly affect the band structure \cite{FlayacBOs1}, and then the confinement energies are pondered by the polariton photonic fraction.
%---

%-------------- 1 --------------

\begin{figure}[ht]
\includegraphics[width=0.5\textwidth,clip]{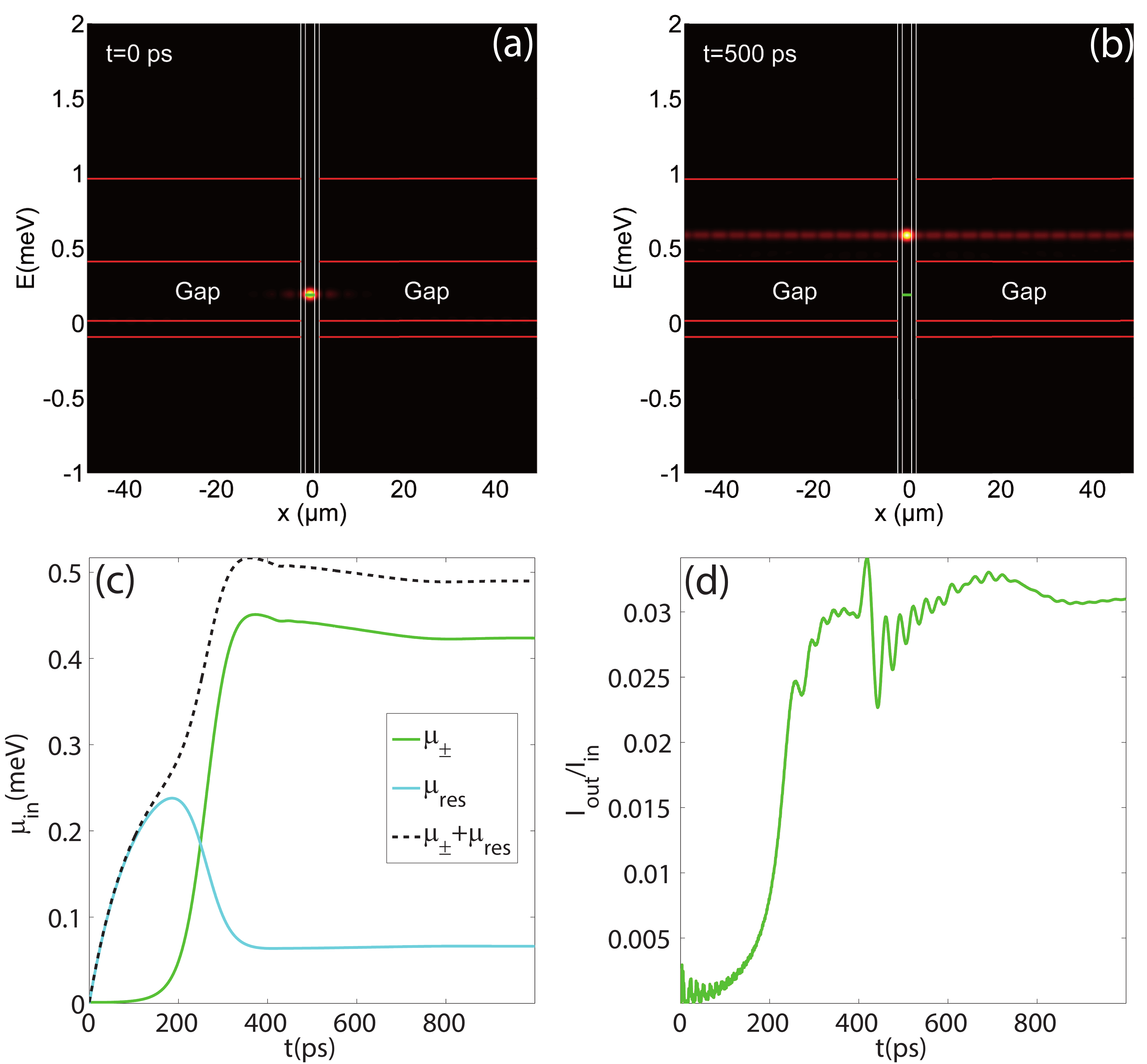}\\
\caption{Illustration of a symmetric optical gate. (a,b) Pattern of the time resolved energy versus space. The colormap shows the total polariton emission intensity, $|\psi_+|^2+|\psi_-|^2$; vertical lines display the barriers; horizontal lines highlight the band edges; the green line marks the initial localized state position. (a) at $t=0$ no transmission occurs when the localized state energy corresponds to the gap of the guides. (b) The reservoir lifts the localized state at the energy of an allowed band thus driving the transmission into the guides. (c) Reservoir (light-blue line) and the condensate (green line) input blueshifts. Dashed black line shows the total blueshift of the localized state. (d) Corresponding normalized output intensities, $I_{\textrm{out}}$, measured in the guides.}
\label{Fig2}
\end{figure}

Let us, first, investigate the case of a fully symmetric sample shown in Fig.~\ref{Fig1}(a) and embodying an optical gate. The results of numerical modelling for this setup are presented in Fig.~\ref{Fig2}. The exciton reservoir is fed by a cw nonresonant pump $P_R(x)$ focused on the inter-barrier region. The localized states are being blueshifted by the energy $\mu_{\rm{res}}$ until the condensation threshold is passed when the polaritons form a condensate. The latter has a linear polarization given that $|\alpha_2|<\alpha_1$ \cite{LinearPolarization} meaning that $|\psi_+|^2=|\psi_-|^2$. The ingoing beam can now be transmitted to the guides provided that its energy is resonant with an allowed energy band.
Panels (a) and (b) are the figure of merit of the effect. They represent snapshots of the time resolved emission as a function of energy and position. As shown in panel (a), initially (at $t=0$ ps) the weakly populated localized state lies at the energy of the gap and therefore it behaves as a defect state of the periodic potential: no transmission occurs into the guides. After 200 ps, the condensate forms leading to the signal transfer through the guides at $t=500$ ps, as shown in panel (b).
Panel (c) accounts for the dependence of the condensate and reservoir blueshifts on time. The corresponding evolution of the total normalized polariton output intensity, $I_{\rm{out}}=|\psi_+|^2+|\psi_-|^2$, is presented in panel (d), calculated at a distance 5 $\mu$m away from the barriers.

In the configuration described above, the system behaves additionally as a \emph{frequency elevator}, in which the output frequency is mapped by the tunable energy of the condensate: $E_{\pm}^0+\mu_{\rm{res}}+\mu_\pm$, where $E_{\pm}^0$ is the inital energy of lowest confined states [green line]. The output minibands embody "floors", accessible to the input beam. A clear visual demonstration of the concept can be found in a Supplemental movie~\cite{Supplemental}: the gate beam is turned on and off periodically corroborating the reproducibility of the theoretical experiment.

A strong asset is the following: the pumping power does not have to be finely tuned. Indeed, as soon as a band is matched, a fraction of the interaction energy is lost through the guides, thus the negative feedback works, on the one hand. On the other hand, the state naturally stabilizes at its position in virtue of the positive feedback from the reservoir feeding.

%-------------- 2 --------------

\begin{figure}[ht]
\includegraphics[width=0.5\textwidth,clip]{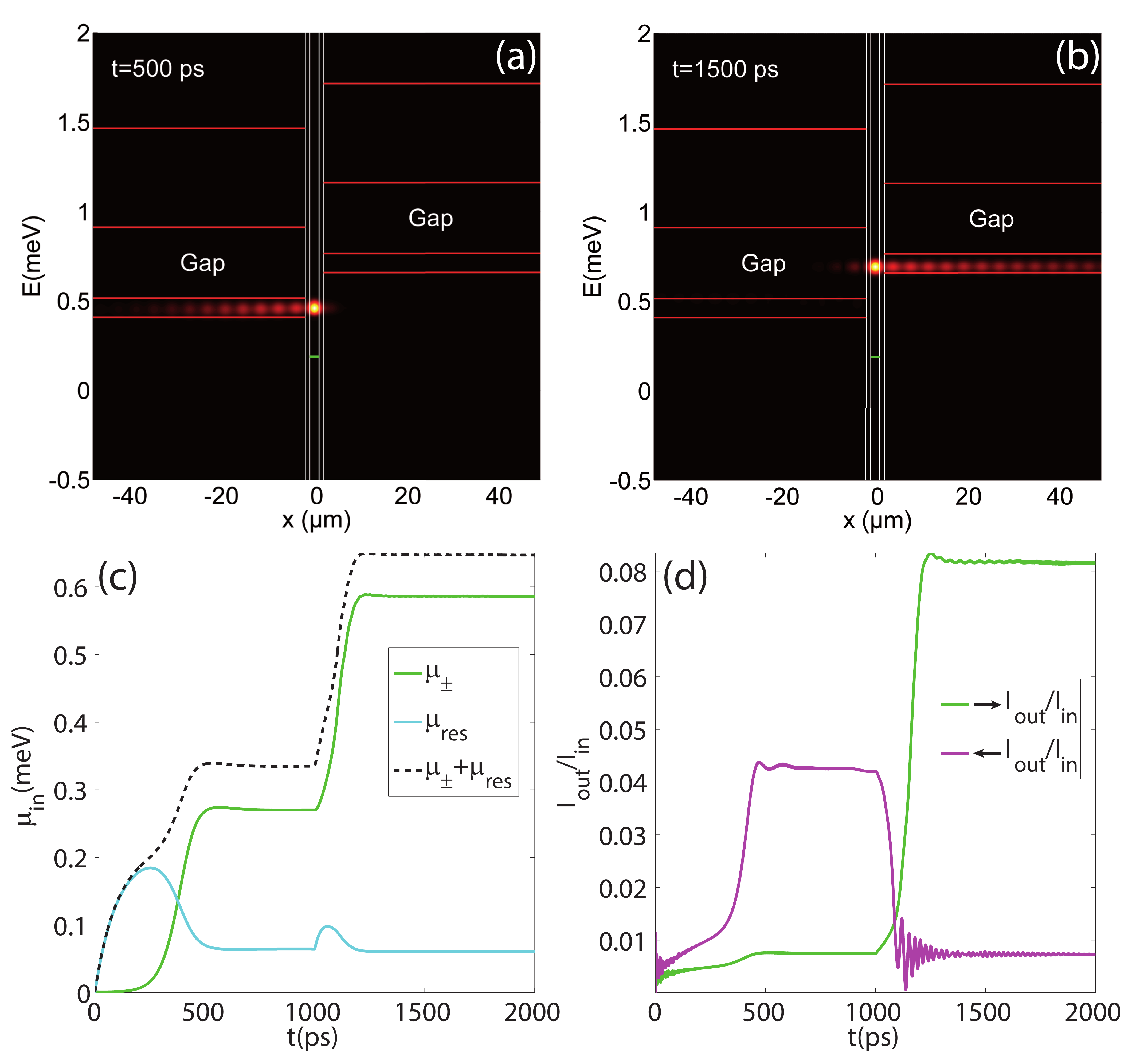}\\
\caption{Asymmetric sample [see Fig.\ref{Fig1}(c)].  The left and right hand guides have a bands mismatch allowing for the beam routing [compare to Fig.\ref{Fig2}]. Varying the pump intensity, we can trace the signal beam to either (a) left or (b) right hand channel. (c) Input blueshifts (the pump power is increased at $t=1000$ ps.) (d) Purple line and green line are the normalized output intensities of the two branches.}
\label{Fig3}
\end{figure}

Second, we consider the case of an asymmetric sample, as shown in Figs.~\ref{Fig1}(c,d) that behaves as an optical router. From the side of geometry, it is assumed that while the periodic potential preserves its shape at both sides, the confinement is set to be stronger on the right hand side of the sample. It requires $\ell_1$, $\ell_2$ and $\ell_2-\ell_1$ in Eq.(\ref{l1l2}) to be smaller than in the left guide. Thus, the right hand periodic potential is simply blueshifted with respect to the left hand one. This asymmetry introduces an energy band mismatch between the two output channels [see horizontal lines in panels Fig.\ref{Fig3}(a,b)].

In that case it turns out possible to route the beam monitoring it by the energy of the localized state. The corresponding results are shown in Fig.\ref{Fig3}. In order to drive the localized state in resonance with the first miniband (from the left hand side) in the time interval 500-1000 ps, and with the second miniband (from the right hand side) from 1200 to 2000 ps, we have used a gradually adjusted pump power [plateaux of the panel (c)]. Panel (d) illustrates the dependence of both the left hand (purple) and right hand (green line) output intensities, manifesting a perfect controllability of the structure~\cite{Supplemental}. It should be mentioned, that the sketch can be easily upgraded by introduction of extra outputs having different band mismatch or bound to a single input. Furthermore, using a sequence of such routers it becomes possible to cascade the condensation within multiple barriers, thus allowing to engineer complex networks.

%-------------- 3 --------------

\begin{figure}[ht]
\includegraphics[width=0.5\textwidth,clip]{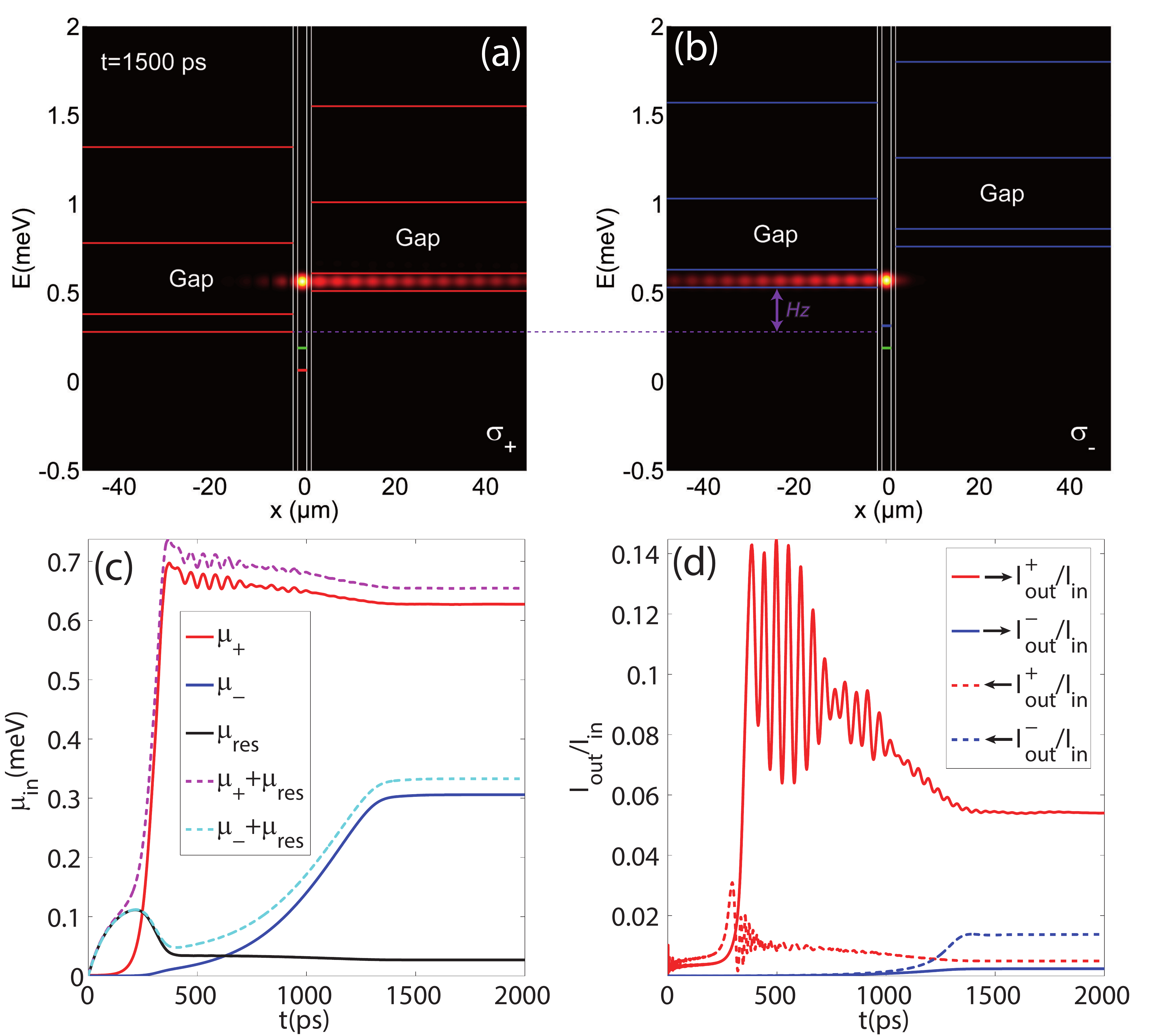}\\
\caption{Illustration of the spin selectivity effect under an applied magnetic field. (a) shows the $\sigma_+$ and (b) the $\sigma_-$ components tracing. Red (blue) line marks the initial $\sigma_+$ ($\sigma_-$) Zeeman split state positions; (c) polarization resolved (c) input and (d) output. In (c,d) the blueshifts experienced by each component are separated (see legend).}
\label{Fig4}
\end{figure}

Third, let us show how the polarization selectivity can come into play. If a magnetic field is applied along the growth axis of the asymmetric sample. It results in a Zeeman splitting $H_z$ between the $\sigma_+$ and $\sigma_-$ spin components [see red and blue lines in Fig.\ref{Fig4}(a,b)]. The spin degeneracy of the localized states is lifted and, consequently, the condensate forms on the lowest energy state that is e.g. $\sigma_+$ polarized for $H_z>0$. While both the spin states are equally blueshifted by the reservoir, only the $\sigma_+$ component suffers the additional polariton blueshift $\mu_+=\alpha_1|\psi_+|^2$, accompanying the condensate onset. Further, if the condensate population is large enough to satisfy the inequality $\mu_+>H_z$, the energy $E_+$ of the $\sigma_+$ mode becomes larger than the energy $E_-$ of the $\sigma_-$ mode. And then, the condensate starts to populate the lowest $\sigma_-$ mode in turn. Thus, the system moves towards a steady state, characterized by the balance condition: $E_+=E_-$.

Meanwhile, in the left and right hand waveguides, the energy bands are equally affected by the Zeeman splitting, thus introducing, additionally, a spin-dependent mismatch between the $\sigma_\pm$ bands for a given output [see red and blue lines in Fig.\ref{Fig4}(a,b)]. This is the key mechanism of the spin selectivity. The value of the magnetic field can be chosen such that to have an exact matching between the right hand $\sigma_-$ and the left hand $\sigma_+$ bands ($H_z=0.25$ meV in our case).

All these preliminary dispositions allow to obtain the following operation mode: the $\sigma_+$ polarized output in the right hand guide [panel (a)] and the $\sigma_-$ output in the left hand guide [panel (b)] \cite{Supplemental}. This is highlighted in panels (c) and (d), on which the dynamics of spins of the input and both the outputs are resolved and separately demonstrated [see legends]. From panel (c) it can be observed, that while in the steady state $E_+=E_-$, the populations of the modes differ since $\mu_+>\mu_-$ as a result of the initial splitting.
Noteworthy, while the lateral etching prevents tuning the energy bands on demand, this constrain can be relaxed here vulnerable to the magnetic field action.

Finally, it should be noted that the effects discussed herein could be reproduced using a resonant gate beam instead of a nonresonant pumping, which has been considered. The switching times become significantly reduced, since one can benefit from the bistability and multistability effects. Unfortunately, the effect turns out extremely unstable due to sensitivity, and the tuning of the output becomes a challenging assignment. Moreover, the required pump powers in this latter case should be much stronger, and a polarized input is required, thus such an alternative appear to be debatable.

%-----------------------------------------------------------------------------------

%*****************************************
%*****************************************

%\emph{Conclusion.---} In summary, we have proposed a set polariton-based all-optical effects capitalizing on the nonlinear excitation of a localized state monitoring a coherent signal tunneling in periodically modulated waveguides. It has been demonstrated, that the symmetric sketch can act as an optical switch and frequency elevator, while introduction of a structural asymmetry between the guides allows to utilize the sample as a quantum router. The application of a magnetic field infers the spin selectivity.

The authors thank E. Ostrovskaya and T. Liew for useful discussions. H. F. acknowledges financial support from NCCR Quantum Photonics (NCCR QP), research instrument of the Swiss National Science Foundation (SNSF).
I. G. S. was supported by the FP7 IRSES projects "SPINMET", "POLAPHEN" and the Eimskip Fund.

%_________________________________________
%--------------------------------------------------------------


\begin{thebibliography}{99}

\bibitem{GaoPRB2013} T. Gao, P. S. Eldridge, T. C. H. Liew, S. I. Tsintzos, G. Stavrinidis, G. Deligeorgis, Z. Hatzopoulos, and P. G. Savvidis, Phys. Rev. B \textbf{85}, 235102 (2012).

\bibitem{BallariniNatCom2013} D. Ballarini, M. De Giorgi, E. Cancellieri, R. Houdr\'{e}, E. Giacobino, R. Cingolani, A. Bramati, G. Gigli and, D. Sanvitto, Nat. Commun. \textbf{4}, 1778 (2013).

\bibitem{THzPRL1} T. C. H. Liew, M. M. Glazov, K. V. Kavokin, I. A. Shelykh, M. A. Kaliteevski, A. V. Kavokin Phys. Rev. Lett. \textbf{110}, 047402 (2011);

\bibitem{THzPRL2} I. G. Savenko, I. A. Shelykh, M. A. Kaliteevski, Phys. Rev. Lett. \textbf{107}, 027401 (2011).

\bibitem{Imamoglu} A. Imamoglu and J. R. Ram, Phys. Lett. A \textbf{214}, 193, (1996).

\bibitem{LiewPRL2008} T. C. H. Liew, A. V. Kavokin, and I. A. Shelykh, Phys. Rev. Lett. \textbf{101}, 016402 (2008).

\bibitem{PavlovicPRL2009} I. A. Shelykh, G. Pavlovic, D. D. Solnyshkov, G. Malpuech, Phys. Rev. Lett. \textbf{102}, 046407 (2009).

\bibitem{JohnePRB2010} I. A. Shelykh, R. Johne, D. D. Solnyshkov, and G. and G. Malpuech, Phys. Rev. B \textbf{82}, 153303 (2010).

\bibitem{EOArX2013} T. Espinosa-Ortega and T. C. H. Liew, Phys. Rev. B \textbf{87}, 195305 (2013).

\bibitem{Racquet} C. Sturm, D. Tanese, H. S. Nguyen, H. Flayac, E. Gallopin, A. Lemaître, I. Sagnes, D. Solnyshkov, A. Amo, G. Malpuech, and J. Bloch, arXiv:1303.1649 (2013).

\bibitem{PolaritonDevices} T. C. H. Liew, I. A. Shelykh, G. Malpuech, Polaritonic devices, Physica E \textbf{43} (9), 1543-1568 (2011).

\bibitem{KasprzakNature} J. Kasprzak, M. Richard, S. Kundermann, A. Baas, P. Jeambrun, J. M. J. Keeling, F. M. Marchetti, M. H. Szy- manska, R. Andre, J. L. Staehli, V. Savona, P. B. Lit- tlewood, B. Deveaud and Le Si Dang, Nature \textbf{443}, 409 (2006).

\bibitem{AmoNature} A. Amo, D. Sanvitto, F. P. Laussy, D. Ballarini, E. del Valle, M. D. Martin, A. Lemaitre, J. Bloch, D. N. Krizhanovskii, M. S. Skolnick, C. Tejedor and L. Vina, Nature \textbf{457}, 291 (2009)

\bibitem{LagoudakisJosephson} K. G. Lagoudakis, B. Pietka, M. Wouters, R. Andre and B. Deveaud- Pledran, Phys. Rev. Lett. \textbf{105}, 120403 (2010)

\bibitem{AmoJosephson} M. Abbarchi,	 A. Amo, V. G. Sala, D. D. Solnyshkov, H. Flayac, L. Ferrier, I. Sagnes, E. Galopin, A. Lema\^{\i}tre, G. Malpuech, and J. Bloch, Nature Physics \textbf{9}, 275–279 (2013).

\bibitem{LaussySupercond} F.P. Laussy, A.V. Kavokin, I.A. Shelykh, Phys. Rev. Lett. \textbf{104}, 106402 (2010)

%\bibitem{ShelykhJosephson} I. A. Shelykh, D. D. Solnyshkov, G. Pavlovic, and G. Malpuech, Phys. Rev. B \textbf{78}, 041302 (2008).

\bibitem{ReviewSpin} I. A. Shelykh, A. V. Kavokin, Y. G. Rubo, T. C. H. Liew. and G. Malpuech, Semicond. Sci. Technol. \textbf{25}, 013001 (2010).

\bibitem{Confinement} A. T. Hammack, M. Griswold, L. V. Butov, L. E. Smallwood, A. L. Ivanov, and A. C. Gossard, Phys. Rev. Lett. \textbf{96}, 227402 (2006); R. B. Balili, D. W. Snoke, L. Pfeiffer, and K. West, Appl. Phys. Lett. \textbf{88}, 031110. (2006); O. El Daif, A. Baas, T. Guillet, J.-P. Brantut, R. Idrissi Kaitouni, J. L. Staehli1, F. Morier-Genoud, and B. Deveaud, Appl. Phys. Lett. \textbf{88}, 061105 (2006); R. I. Kaitouni, O. El Daif, A. Baas, M. Richard, T. Paraiso, P. Lugan, T. Guillet, F. Morier-Genoud, J. D. Ganiere, J. L. Staehli, V. Savona, and B. Deveaud, Phys. Rev. B \textbf{74}, 155311 (2006); M. M. Kaliteevskii, S. Brand, R. Abram, I. Iorsh, A. Kavokin, and I. Shelykh,  Appl. Phys. Lett. \textbf{95}, 251108 (2009)

\bibitem{WertzNature} E. Wertz, L. Ferrier, D. Solnyshkov, R. Johne, D. Sanvitto, A. Lemaitre, I. Sagnes, R. Grousson, A. V. Kavokin, P. Senellart, G. Malpuech, and J. Bloch, Nature Physics \textbf{6}, 860 (2010)

\bibitem{LiewCircuit} T. C. H. Liew, A. V. Kavokin, T. Ostatnicky, M. Kaliteevski, I. A. Shelykh, and R. A. Abram  Phys. Rev. B \textbf{82}, 033302 (2010)

\bibitem{TaneseGS} D. Tanese, H. Flayac, D. Solnyshkov, A. Amo,	A. Lema\^{\i}tre, E. Galopin, R. Braive, P. Senellart, I. Sagnes, G. Malpuech and J. Bloch, Nat. Commun. \textbf{4}, 1749 (2013).

\bibitem{Sollner1983} T. C. L. G. Sollner, W. D. Goodhue, P. E. Tannenwald, C. D. Parker, and D.D. Peck, Appl. Phys. Lett. \textbf{43}, 588 (1983).

\bibitem{Mizuta1995} H. Mizuta and T. Tanoue, The Physics and Applications of Resonant Tunneling Diodes (Cambridge University Press, New York, 1995).

\bibitem{Pasquier1993} C. Pasquier, U. Meirav, F. Williams, D. Glattli, Y. Jin, and B. Etienne, Phys. Rev. Lett. \textbf{70}, 69 (1993).

\bibitem{Joyez1994} P. Joyez, P. Lafarge, A. Filipe, D. Esteve, and M. H. Devoret, Phys. Rev. Lett. \textbf{72}, 2458 (1994).

\bibitem{Carusotto2000} I. Carusotto and G. C. LaRocca, Phys. Rev. Lett. \textbf{84}, 399 (2000).

\bibitem{Paul2005} T. Paul, K. Richter, and P. Schlagheck, Phys. Rev. Lett. \textbf{94}, 020404 (2005).

\bibitem{AmoPRB2010} A. Amo, S. Pigeon, C. Adrados, R. Houdr\'e, E. Giacobino, C. Ciuti, and A. Bramati, Phys. Rev. B \textbf{82}, 081301 (2010).

\bibitem{TosiNature2012} G. Tosi, G. Christmann, N. G. Berloff, P. Tsotsis, T. Gao, Z. Hatzopoulos, P. G. Savvidis, and J. J. Baumberg, Nat. Phys. \textbf{8}, 190 (2012).

\bibitem{LaiNature2007} C. W. Lai, N. Y. Kim, S. Utsunomiya, G. Roumpos, H. Deng, M. D. Fraser, T. Byrnes, P. Recher, N. Kumada, T. Fujisawa et al., Nature (London) \textbf{450}, 529 (2007).

\bibitem{BlochSM1997} J. Bloch, R. Planel, V. Thierry-Mieg, J. M. Ge ́ rard, D. Barrier, J.Y. Marzin, and E. Costard, Superlattices Microstruct. \textbf{22}, 371 (1997).

\bibitem{ParaisoPRB2009} T. K. Para\:iso, D. Sarchi, G. Nardin, R. Cerna, Y. Leger, B. Pietka, M. Richard, O. El Da\:f, F. Morier-Genoud, V. Savona  et al., Phys. Rev. B \textbf{79}, 045319 (2009).

\bibitem{NguyenPRL2013} H. S. Nguyen, D. Vishnevsky, C. Sturm, D. Tanese, D. Solnyshkov, E. Galopin, A. Lema\'itre, I. Sagnes, A. Amo, G. Malpuech, and J. Bloch, Phys. Rev. Lett. \textbf{110}, 236601 (2013).

\bibitem{WCModel} M. Wouters and I. Carusotto, Phys. Rev. Lett. \textbf{99}, 140402 (2007).

\bibitem{LiewTransistor} I. G. Savenko, T. C. H. Liew, and I. A. Shelykh, Phys. Rev. Lett. \textbf{110}, 127402 (2013).

\bibitem{SavenkoPRL2013} I. G. Savenko, E. B. Magnusson, and I. A. Shelykh, Phys. Rev. B \textbf{83}, 165316 (2011).

%\bibitem{YamamotoNature2007} C. W. Lai, N. Y. Kim, S. Utsunomiya, G. Roumpos, H. Deng, M. D. Fraser, T. Byrnes, P. Recher, N. Kumada, T. Fujisawa, and Y. Yamamoto, Nature \textbf{450}, 529-532 (2007).

\bibitem{FlayacBOs1} H. Flayac, D. D. Solnyshkov, and G. Malpuech, Phys. Rev. B \textbf{83}, 045412 (2011).

\bibitem{FlayacBOs2} H. Flayac, D. D. Solnyshkov, and G. Malpuech Phys. Rev. B \textbf{84}, 125314 (2011).

\bibitem{SAW} E. A. Cerda-Mendez, D. N. Krizhanovskii, M.Wouters, R. Bradley, K. Biermann, K. Guda, R. Hey, P. V. Santos, D. Sarkar, and M. S. Skolnick, Phys. Rev. Lett. \textbf{105}, 116402 (2010).

\bibitem{LinearPolarization} J. Kasprzak, R. André, Le Si Dang, I. A. Shelykh, A. V. Kavokin, Yuri G. Rubo, K. V. Kavokin, and G. Malpuech, Phys. Rev. B \textbf{75}, 045326 (2007).

\bibitem{Supplemental} See supplementary material at [URL will be inserted by AIP] for a movie.
%%%

\end{thebibliography}
\end{document}